\newcommand{\newc}{\newcommand}

\newc{\Lumi}{{\cal L}}

\newc{\ra}{\rightarrow}
\newc{\Ra}{\Rightarrow}

\newc{\pom}  {I\hspace{-0.2em}P}

\newc{\rpv}{{\not \!\! R_p}}
\newc{\rpvm}{{\not  R_p}}

\newc{\gsim}{{\stackrel{>}{\sim}}}
\newc{\lsim}{{\stackrel{<}{\sim}}}
\newc{\sleq} {\raisebox{-.6ex}{${\textstyle\stackrel{<}{\sim}}$}}
\newc{\sgeq} {\raisebox{-.6ex}{${\textstyle\stackrel{>}{\sim}}$}}

\def\3{\ss}

\newc{\ETJ}{E^{{\rm jet}}_T}
\newc{\PTJ}{p^{{\rm jet}}_T}

\def\xgo{x_\gamma^{\rm OBS}}

\def\pt{p_T}

\def\kt{K_\perp~}

\def\ee{e^+ e^-}

\def\q2{{\rm Q}^2}
\def\p2{{\rm P}^2}
\def\gev{{\rm GeV}}
\def\gevsq{{\rm GeV}^2}

\def\F2g{F_2^\gamma}
\def\f2c{F_2^{\rm c\bar{c}}}

\def\d0{D^{0}}

\def\begr{\begin{flushright}}
\def\endr{\end{flushright}}

\def\begl{\begin{flushleft}}
\def\endl{\end{flushleft}}

\def\as{\alpha_s}

\def\jimmy{{\sc Jimmy}}
\def\herwig{{\sc Herwig}}
\def\pythia{{\sc Pythia}}
\def\mcnlo{{\sc MC@NLO}}
\def\sherpa{{\sc Sherpa}}
\def\cascade{{\sc Cascade}}

\documentclass{ws-procs10x7}

\begin{document}

\title{Quantum Chromodynamics at Colliders}

\author{J. M. Butterworth}

\address{Department of Physics \& Astronomy, University College
    London, WC1E 6BT, London, UK\\E-mail: J.Butterworth@ucl.ac.uk}

\twocolumn[\maketitle\abstract{ QCD is the accepted (that is, the
effective) theory of the strong interaction; studies at colliders are
no longer designed to establish this. Such studies can now be divided
into two categories. The first involves the identification of
observables which can be both measured and predicted at the level of a
few percent. Such studies parallel those of the electroweak sector
over the past fifteen years, and deviations from expectations would be
a sign of new physics. These observables provide a firm ``place to
stand'' from which to extend our understanding. This links to the
second category of study, where one deliberately moves to regions
in which the usual theoretical tools fail; here new approximations in QCD
are developed to increase our portfolio of understood processes, and
hence our sensitivity to new physics.  Recent progress in both these
aspects of QCD at colliders is discussed.  }]

\section{The Data and the Experiments}

QCD studies at colliders involve measurements of the hadronic final
state in $\ee$, lepton-hadron and hadron-hadron collisions. The lepton
colliders also allow the study of effective photon-photon,
lepton-photon and photon-hadron collisions, due to the almost-on-shell
photon beam which accompanies lepton beams. In collisions involving
these photons, the photon may participate directly in the hard
process, or it may act as a source of partons much like a
hadron. Together, this array of different colliding beams provides us
with many data and rich opportunities to learn from cross-comparison
between experiments.

Data presented at this meeting include precise measurements of a great
number of properties of the final state, and these measurement are
used to demonstrate and improve our understanding of the physics. With
the confidence that this is understood, it then becomes possible to
infer, from an increasing number of measurements, information about
the initial state; that is, quarks and gluons in their natural
habitat inside hadrons. This in turn enables us to predict effects at
future colliders, particularly the Large Hadron Collider under
construction at CERN~\cite{lhc}.

In sections 2-5, the final state measurements are discussed. In the
subsequent section, some experimental advances in the current
knowledge of parton densities within the
Dokshitzer-Gribov-Lipatov-Altarelli-Parisi (DGLAP) paradigm are
presented. Following that, some measurements in regions of phase space
where DGLAP evolution is not applicable are discussed. This includes
low $x$ and diffractive effects, at which point I conclude this
contribution and hand over to the next speaker\cite{gunnar}.

\section{Fragmentation and Hadron Production}

\begin{figure}
\epsfxsize200pt
\figurebox{120pt}{160pt}{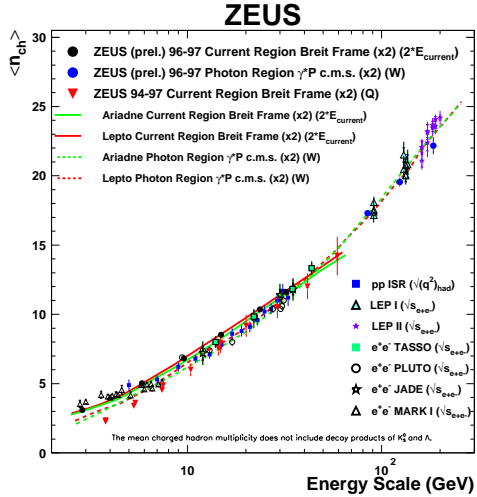}

\vspace{-0.5cm}

\caption{The charged particle multiplicity as a function of energy
scale for a selection of experiments.}
\label{fig:cmult}
\end{figure}

An obvious observable to start with in looking at QCD final states is
the charged particle multiplicity. This has been measured as a
function of the energy scale of the interaction by many experiments. A
summary\cite{zmult} is shown in Fig.~\ref{fig:cmult}. The energy scale
dependence is seen to be universal to within a few percent for
reasonable definitions of the energy scale in $\ee$ and DIS, and the
proton data from ISR also lies close to the same curve. This is well
modelled by the current Monte Carlo (MC) models. The shape is also
described by next-to-leading-order (NLO) QCD (not shown), where local
parton-hadron duality is assumed to give an arbitrary constant
normalisation factor.

To make more precise statements about QCD fragmentation, measurements
can be designed specifically to suit precise calculations. Accurate
calculations for quark and gluon fragmentation exist for hemispheres
of a fragmenting diquark of di-gluon system. In the case of quarks,
this is a natural configuration for comparison with $\ee$ data.
Obtaining a comparable configuration for gluons, however, is more
difficult. In a contribution from OPAL\cite{jboost} the jet boost
algorithm is employed to do this. Precise agreement is observed for
$0.06 < x < 0.8$. Because of this level of agreement, fundamental
parameters of the theory can be extracted with confidence.  An
impressive recent example is the measurement of the ration of the
gluon and quark colour factors, $C_A/C_F = 2.261 \pm 0.014 \pm 0.036
\pm 0.066$, by DELPHI\cite{cacf}, where the first error is statistical, the
second the experimental systematic error and the the third the
theoretical uncertainty. This agrees well with the QCD expectation of
2.25.

\begin{figure}
\epsfxsize200pt
\figurebox{120pt}{160pt}{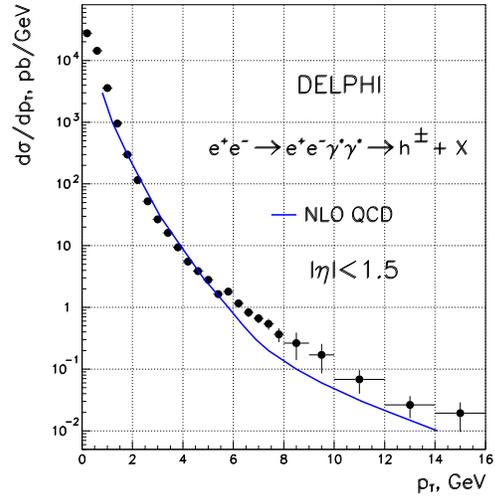}

\epsfxsize200pt
\figurebox{120pt}{160pt}{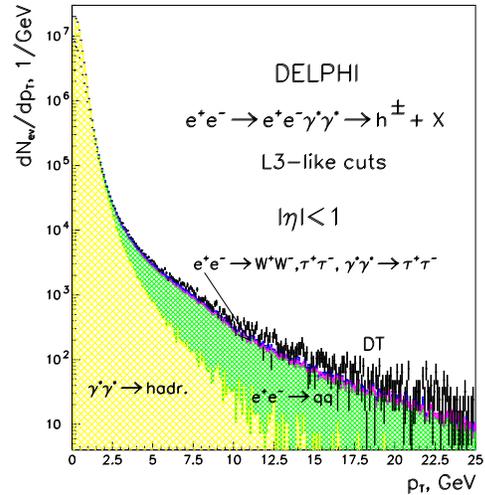}

\vspace{-2.0cm}

\caption{The charged particle cross section in $\gamma\gamma$ collisions 
as a function of particle transverse momentuym ($\pt$) 
as measured by DELPHI. The upper plot is the DELPHI measurement 
of the cross section compared to NLO QCD. The lower plot is the
DELPHI data analysed using cuts close to those used by L3 (see
text).}
\label{fig:dpt}
\end{figure}

One assumption employed in such measurements is that the soft,
hadronization stage can be controlled and seperated from the hard QCD
process. This assumption has been tested in many measurements, and
several new results from HERA\cite{heracfrag} have tested it in the
case of charm quarks.  Here it has been shown that the fraction of
charm quarks fragmenting to the various charmed hadrons is the same
(to within the measurement accuracy of a few \%) in DIS and
photoproduction at HERA as it is in $\ee$ annihilation. Comparisons
between the fragmentation function at HERA, LEP and CLEO also show
qualitative agreement. A fit of the fragmentation function using NLO
calculations would allow a more quantitative statement to be made here,
and would be of great interest; as would more accurate measurements
from HERA II.

The claim is that for some QCD observables the theoretical
understanding is so good that deviations in the data really do mean
new physics. This claim was challenged by two results from the L3
collaboration, where in $\gamma\gamma$ events, both the charged
particle and jet cross sections lie above the NLO QCD prediction, with
a discrepancy which increases as the scale
increases\cite{l3stuff}. This discrepancy seems impossible to
reconcile with QCD; yet the scale is so low ($\pt
\approx 5 \gev$ for the charged hadrons) that some
beyond-the-standard-model explanation seems unlikely.  The charged
particle measurement has been repeated by DELPHI\cite{delphipt},
however, and no such discrepancy is seen (Fig.\ref{fig:dpt} - note
that no theoretical uncertainty is shown). To their great credit,
DELPHI have gone further, solving the puzzle by mimicking the L3
analysis and showing that for the L3 selection cuts there is a large
background from annihilation, which has the correct charactierstics to
explain the discrepancy. This is also shown in Fig.\ref{fig:dpt}; it
is then a victory for some kind of precision QCD. It is tempting to
speculate that the $\ee$ background may also contribute to the excess
seen by L3 in the jet cross section.

\begin{figure}
\epsfxsize200pt
\figurebox{120pt}{160pt}{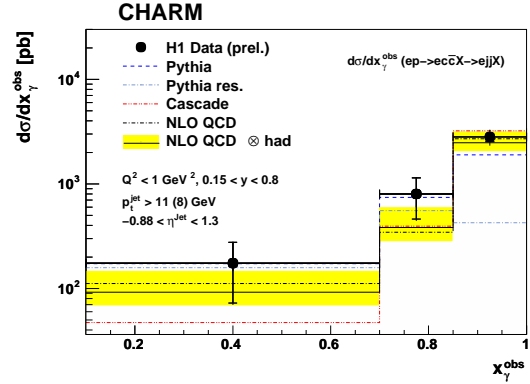}
\vspace{-0.5cm}
\caption{The $\xgo$ distribution in charm photoproduction.}
\label{fig:h1xgc}
\end{figure}

\section{Charm and Beauty Production}

Recent data on fragmentation properties of charm have been briefly
discussed above. The production cross sections for both charm and
bottom quarks also represent an important investigative tool for QCD,
and since bottom in particular is often used as a tag in searches for
new physics, the QCD production mechanism is of particular importance.
An understanding of the production dynamics as well as inclusive rates
is needed. Results continue to be produced from $p\bar{p}$, $ep$ DIS
and photoproduction.

\begin{figure}
\epsfxsize200pt
\figurebox{120pt}{160pt}{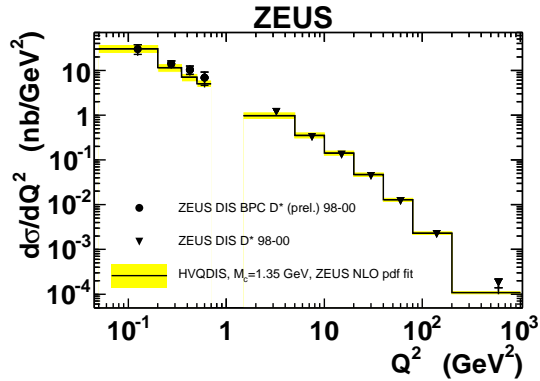}
\vspace{-0.5cm}
\caption{The inclusive charm cross section as a function of $\q2$.}
\label{fig:cq2}
\end{figure}

\subsection{Charm cross sections}

\begin{figure}
\epsfxsize180pt
\figurebox{120pt}{160pt}{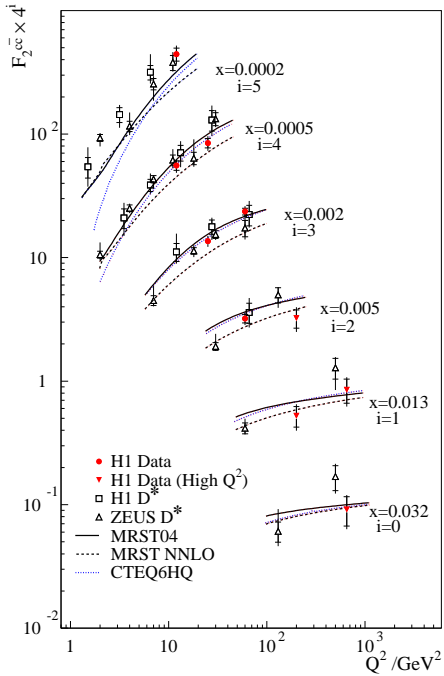}

\vspace{-0.2cm}

\caption{The charm tagged structure function of the proton.}
\label{fig:f2c}
\end{figure}

Photoproduction of charm has been measured using tagged $D^* + $ jets
and via lifetime tagging\cite{h1cbphp,zcphp,review}. Changing the
fraction of the photon's momentum seen in the jets, $\xgo$, from
values near one to lower values allows one to move from so-called
direct processes, dominated by point-like photons, to resolved
processes, in which the photon acts as a source of partons similar to
a hadron. Both regions are well described by NLO QCD calculations
(Fig.\ref{fig:h1xgc}). In addition, the inclusive cross section is
well understood in both the photoproduction and DIS regimes, from
photon virtualities of near zero up to $1000\gevsq$
(Fig.~\ref{fig:cq2}). Expressed as the charm structure function
$\f2c$, the data is already quite precise and is still being
accumulated. Again, NLO QCD describes it well
(Fig.~\ref{fig:f2c})\cite{h1cbf2}.

On a related topic, inelastic $J/\psi$ production, the debate about
colout octet terms is not yet resolved. NLO QCD corrections to the
colour singlet term are very large\cite{review}.

\subsection{Beauty cross sections}

\begin{figure}

\epsfxsize200pt
\figurebox{120pt}{160pt}{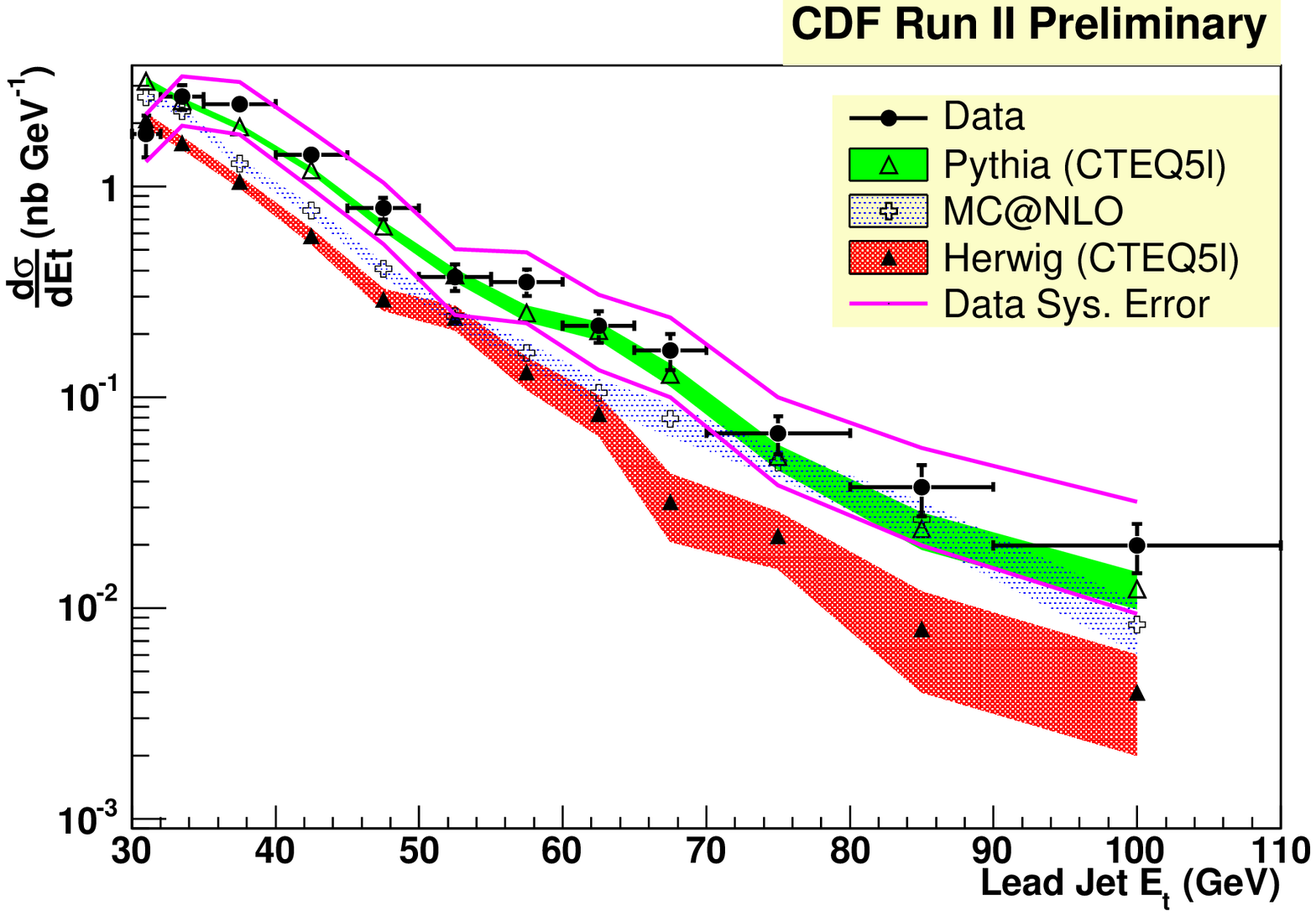}

\epsfxsize200pt
\figurebox{120pt}{160pt}{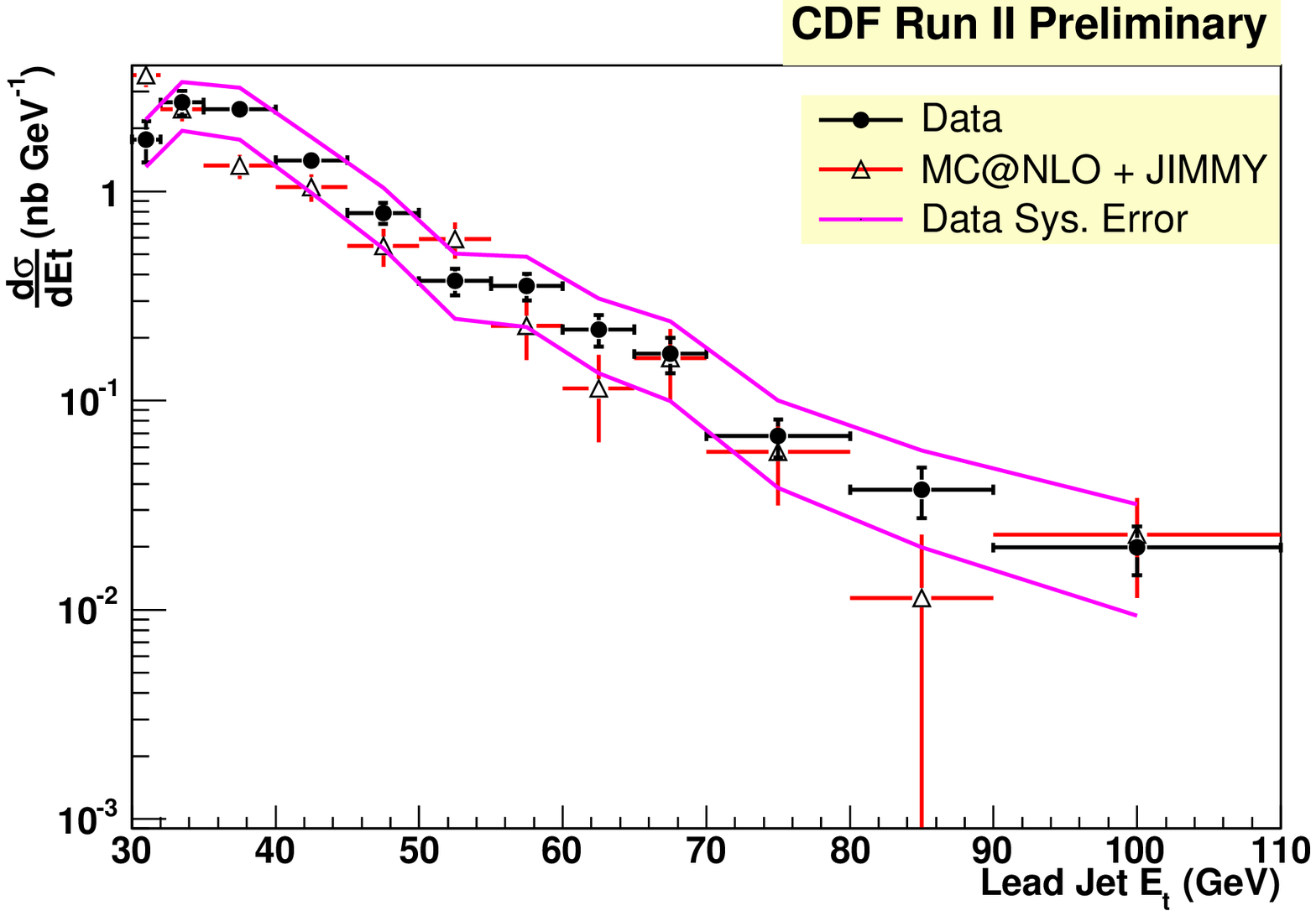}

\vspace{-0.5cm}

\caption{Bottom-quark jet cross sections from CDF II.}
\label{fig:cdfbj}
\end{figure}

Inclusive measurements of bottom-tagged cone dijets from the CDF II
have been measured\cite{bdijets} and compared to \pythia\cite{pythia},
\herwig\cite{herwig} and \mcnlo\cite{mcnlo}
(Fig.~\ref{fig:cdfbj}a). The normalisation of the LO
MCs has a large uncertainty associated with it due to higher
order terms. However, it is significant that \pythia\ describes the
shape of the data very well for $\ETJ > 40
\gev$. \mcnlo\ is in good agreement with the cross section at high
transverse momenta but falls below the data at $\ETJ < 70\gev$. Apart
from the NLO terms, one difference between the two programs is that
\pythia\ includes a multiparton interaction model to describe the
underlying event. Adding such a model to \mcnlo\, in the shape of
\jimmy\cite{jimmy}, leads to good agreement between
\mcnlo\ and the data for $\ETJ > 40 \gev$
(Fig.~\ref{fig:cdfbj}b). 
 
\begin{figure}
\epsfxsize200pt
\figurebox{120pt}{160pt}{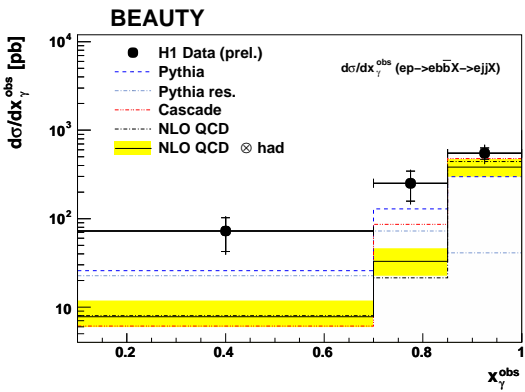}

\vspace{-0.5cm}

\caption{The $\xgo$ distribution in bottom photoproduction.}
\label{fig:xgb}
\end{figure}

There are also measurements from D0 of muon-tagged jets\cite{d0mu},
where within 50\% errors NLO calculations describe the data. At HERA,
DIS and direct photoproduction measurements are reasonably well
described, though there is a tendency for the data to be above the
calculations. This seems particularly pronounced at low $\xgo$ (see
Fig.\ref{fig:xgb}), where it is possible that non-perturbative effects
such as the underlying event may play some role. Precision data from
HERA II will hopefully clarify the situation.

\begin{figure}
\epsfxsize180pt
\figurebox{120pt}{160pt}{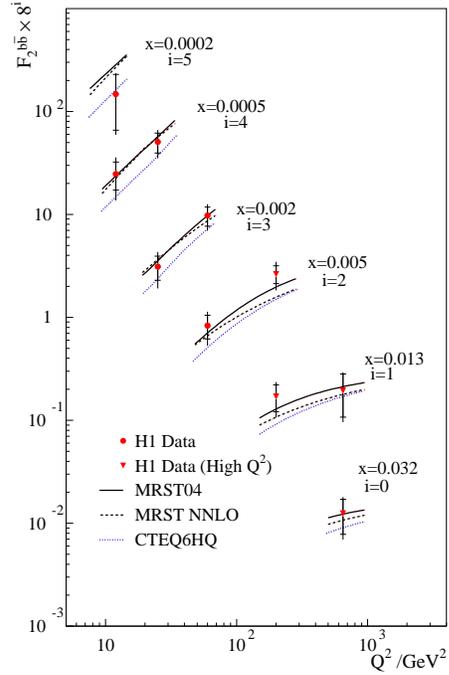}

\vspace{-0.2cm}

\caption{The bottom-quark tagged structure of the proton.}
\label{fig:f2bb}
\end{figure}

Finally, the first measurements of the beauty stucture function
$F_2^{\rm b\bar{b}}$ have now been made~\cite{h1cbf2}, shown
in Fig.~\ref{fig:f2bb}. These lag the similar charm measurements in
statistical precision, but there are many more data to come, and it
will be an important challenge for the theory to describe such
inclusive measurements well.

\subsection{Charm and Bottom production dynamics}

\begin{figure}
\epsfxsize200pt
\figurebox{120pt}{160pt}{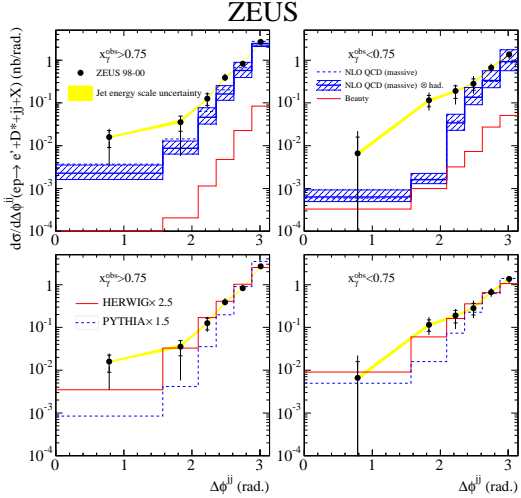}
\caption{The $\Delta\phi$ distribution for charm jets.}
\label{fig:charmdecor}
\end{figure}

The charm statistics at HERA are sufficient that the production
dynamics may be measured. Several measurements already
exist\cite{review}, and there are new measurements now of the
azimuthal correlation of dijets in charm events\cite{zcphp}, as well
as jet shapes for charm jets\cite{h1cbphp}. Both are sensitive to QCD
radiation in these processes. The azimuthal decorrelation is well
described by leading-logarithmic parton shower models for both
resolved and direct photoproduction; NLO calculations for massive
charm quarks (e.g. in which the charm is not an active quark in the
photon or proton) describe the direct case well, but fail to describe
the low-$\xgo$ decorrelation (see Fig.\ref{fig:charmdecor}. The jet
shapes are well described by \pythia's parton showers for high $\xgo$,
but the jets are narrower in the data than in the MC at low $\xgo$.

In the case of beauty, the Tevatron data allow studies of such
properties in bottom quark events. The dijet correlation is reasonably
well described by \mcnlo, but the addition of multiparton interactions
does again improve the agreement. \pythia\ also does a reasonable job.

Finally, a beautiful new measurement of the ratio of bottom- to
light-quark jet rates from DELPHI\cite{bmass} leads to an accurate
measurement of the running b-mass $m_b(Q) = 4.25 \pm 0.11 \gev$ at
threshold.

In summary of this section, it does seem that in general charm and
bottom production are well described by NLO QCD, but that there is a
need to combine state-of-art non-perturbative models with the best
perturbative calculations in order to get this level of
agreement. This is true particularly for measurements in hadronic
collisions spanning a large range in transverse energy.

\section{Jet Structure and Event Shapes}

\begin{figure}
\epsfxsize200pt
\figurebox{120pt}{160pt}{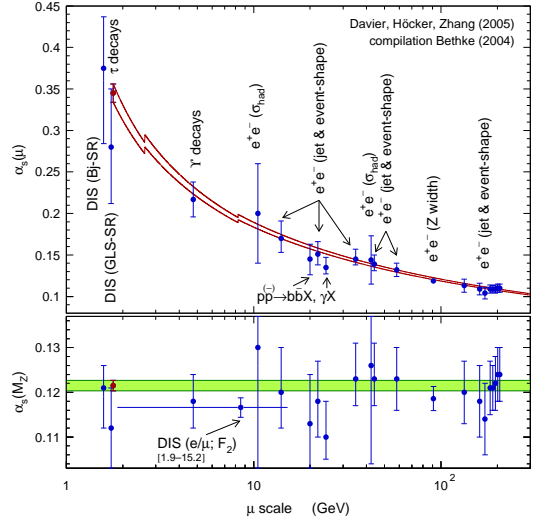}

\vspace{-1.0cm}

\caption{A selection of $\alpha_s$ measurements.}
\label{fig:as}
\end{figure}

Measurements of jet cross sections and event shapes continue to
improve in precision, as do calculations of such properties. This
means that the strong coupling, $\as$, may be extracted from a large
number of final states in many processes. At this conference, new
results from $\ee$ (JADE, OPAL, ALEPH) and $ep$ (H1,
ZEUS) were presented\cite{jetandeshapes,astau}. A particularly interesting
measurement is the ALEPH extraction from $\tau$ decays, shown in
Fig.~\ref{fig:as}, which greatly improves the accuracy at low
scales~\cite{astau}. In general, none of the others is a great leap
forward in itself, but all steadily improve accuracy of the world
average, and build confidence in our understanding of QCD.

\begin{figure}

\epsfxsize160pt
\figurebox{120pt}{160pt}{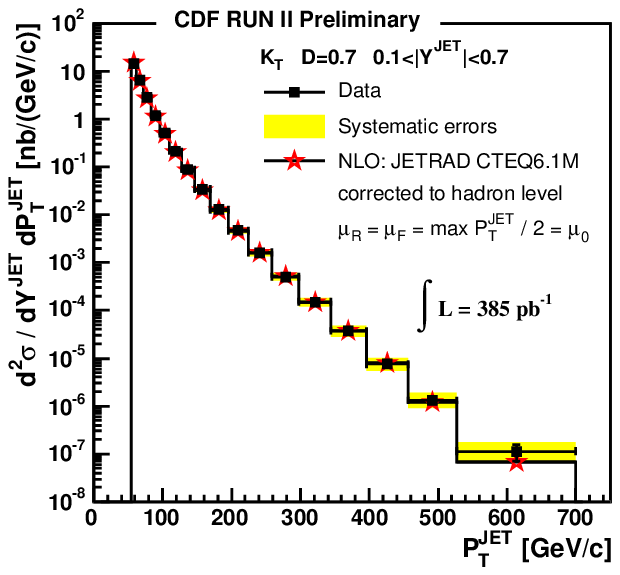}

\epsfxsize160pt
\figurebox{120pt}{160pt}{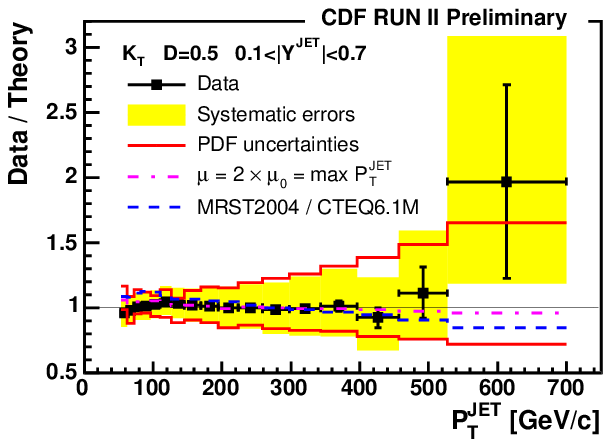}

\epsfxsize160pt
\figurebox{120pt}{160pt}{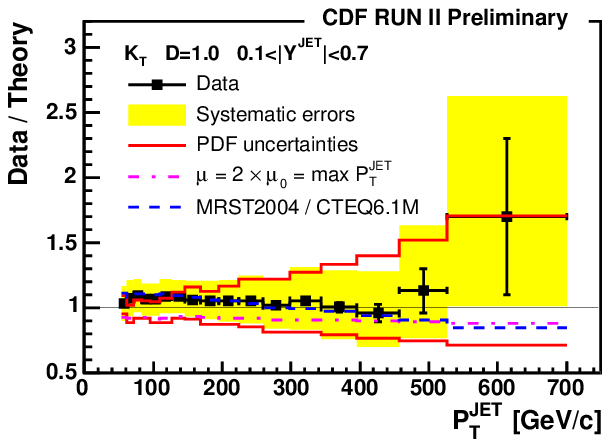}

\caption{CDF inclusive jet measurements using the $\kt$ algorithm. The top plot shows the measured differential cross section $d\sigma/d\PTJ$ compared to NLO QCD for $R=0.7$. The lower two plots show the ratio of data/theory for similar cross sections measured with $R=0.5$ and $R=1.0$.}
\label{fig:incjets}
\end{figure}

Behind this achievement lies an increasing number of well-understood
QCD processes. Perhaps particularly noteworthy this year are the new
inclusive jet measurements from Tevatron Run II and HERA, where the
use of well-controlled jet algorithms and the impressive level of
knowledge of the energy scale and resolution in the experiments means
that the data really lay down a strong challenge for the theoretical
predictions. Some of the CDF II results are shown
Fig.~\ref{fig:incjets}; here the $\kt$ algorithm has been used with
different distance parameters; this is an important technique, in that
any new physics effect seen in such cross sections should be present
for all reasonable choices, whereas the sensitivites to some
non-perturbative effects will vary between different algorithms and
parameters. Another interesting process with new data is prompt photon
production, where both HERA and Tevatron have new
data\cite{prompt,diphoton}. The D0 data in particular now show
impressive agreement with QCD over a wide range of transverse energy.

\section{Production of jets with bosons}

\begin{figure}
\epsfxsize200pt
\figurebox{120pt}{160pt}{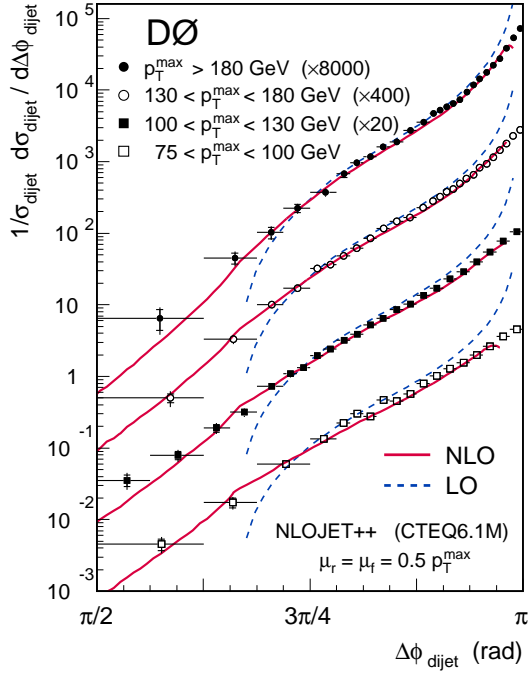}
\caption{Dijet decorrelation from D0.}
\label{fig:d0dij}
\end{figure}

When the LHC starts delivering data, an unprecedented number of $W$
and $Z$ particles will be produced, usually in association with
jets. They feature in many ``standard candle'' cross sections which
will be used to extract parton densities and calibrate the detectors,
as well as in many exotic signatures for new physics. It is imperative
to understand as far as possible equivalent processes at existing
colliders, particularly the Tevatron. The dijet correlation\cite{d0dij} at D0 is
shown in Fig~\ref{fig:d0dij}. It is well described by NLO QCD in the
important wide-angle area where the fixed-order tree-level diagrams
are most significant, and is described by parton shower MC in the low
angle regions, as expected. Importantly, the \sherpa\ program matches
these two types of calculations and describes the whole shape
well\cite{sherpa}.

\begin{figure}
\epsfxsize200pt
\figurebox{120pt}{160pt}{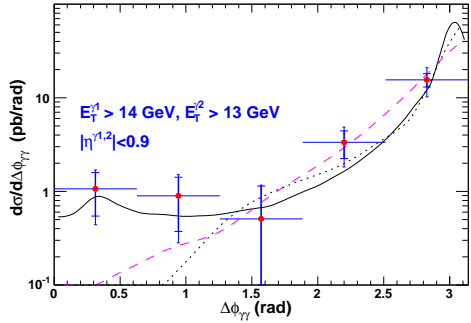}
\caption{Diphoton decorrelation from CDF. The points are the data, the solid line is the DIPHOX calculation and the dashed is RESBOS (see text).}
\label{fig:diph}
\end{figure}

A related cross section is the diphoton decorrelation, measured by
CDF\cite{diphoton}, shown in Fig.~\ref{fig:diph}. The angle between
the two photons is well described by NLO QCD as contained in the
DIPHOX\cite{diphox} program. The RESBOS\cite{resbos} calculation does
not include NLO fragmentation contributions and falls below the data
at high angles.

\begin{figure}

\epsfxsize180pt
\figurebox{120pt}{160pt}{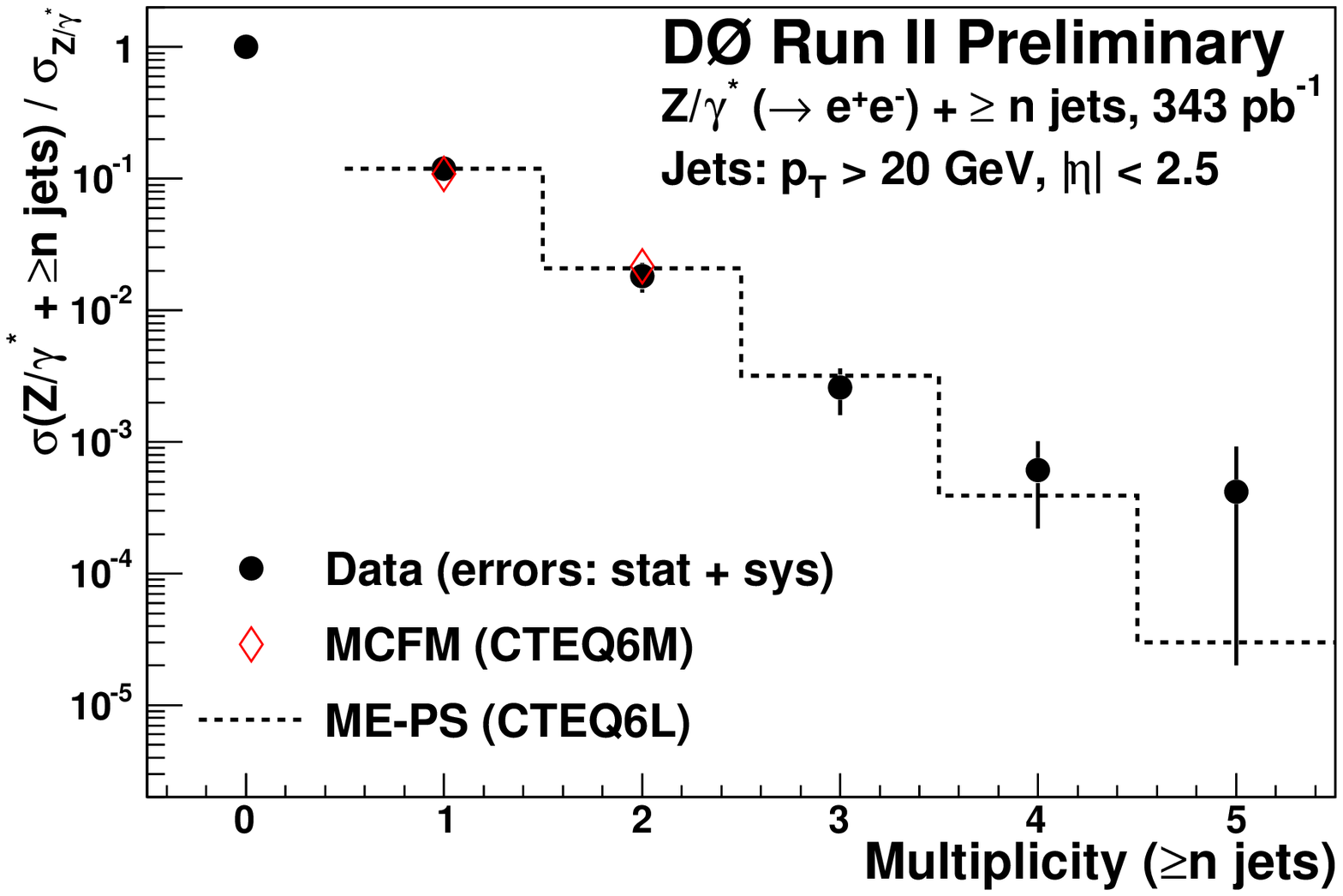}

\epsfxsize180pt
\figurebox{120pt}{160pt}{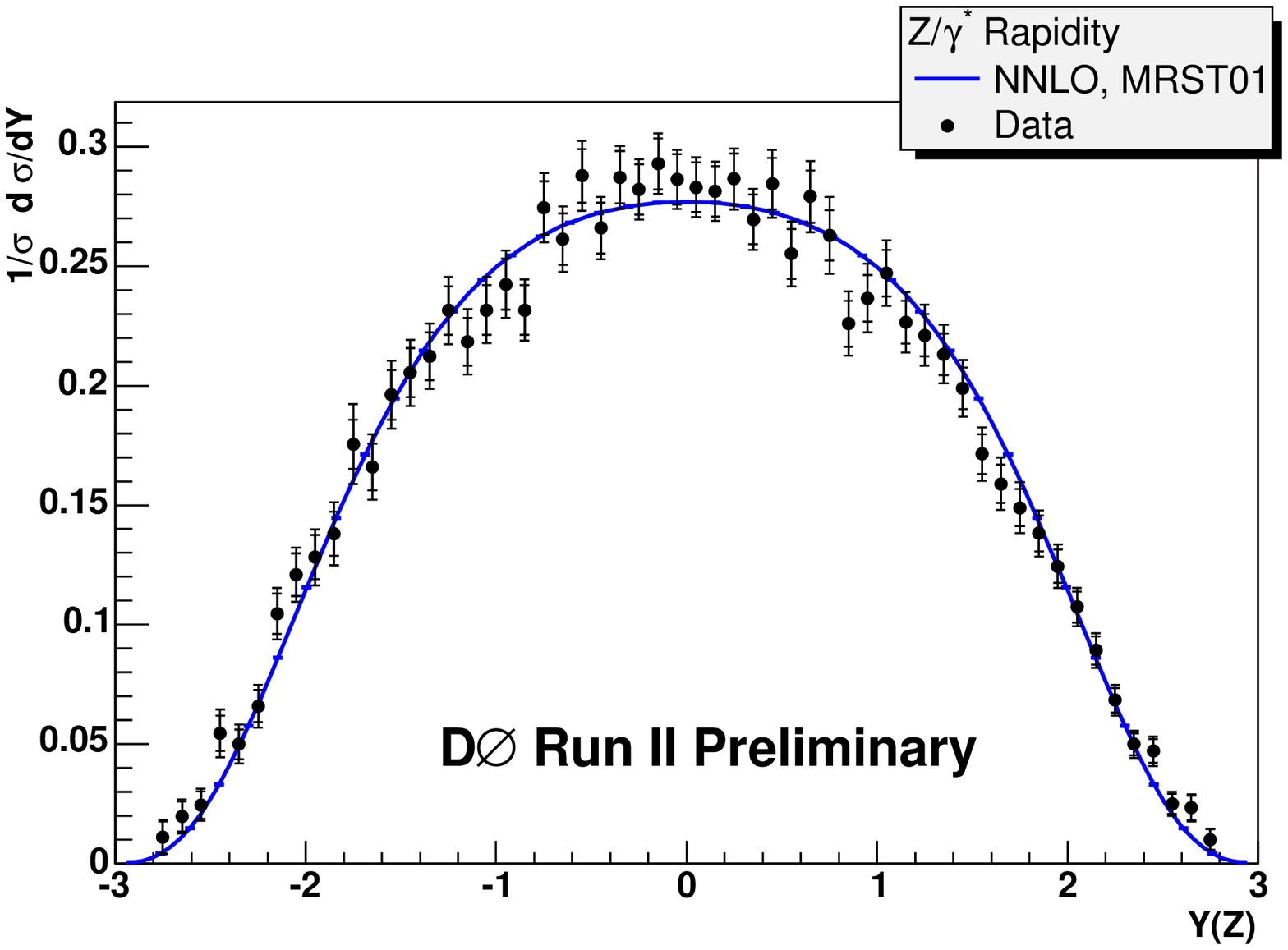}

\caption{The $N$-jet rate in $Z$ events, compared to MC calculation (upper plot) and the $Z$ rapidity distribution (lower plot).}
\label{fig:zs}
\end{figure}

Run II measurements of $Z$ cross sections are now coming out, and both
the inculsive $Z$ rapidity\cite{zrap} and the $N$-jet rate in $Z$
events\cite{znjet} are in good agreement with NLO QCD
(Fig.\ref{fig:zs}).

\section{Parton Densities}

There has been major theoretical progress in this area, as discussed
in the previous contribution\cite{gavin}. There have also been some
notable experimental advances, which are discussed below.


\subsection{High $x$}

\begin{figure}
\epsfxsize160pt
\figurebox{120pt}{170pt}{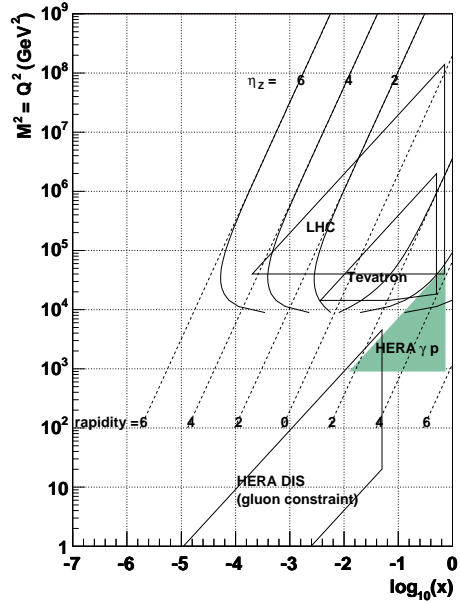}
\vspace{0.1cm}
\caption{The parton kinematics at LHC and elsewhere. The curved lines show the
region where $Z+$jet production might be used to constrain the gluon
at the LHC.The HERA and Tevatron regions shown are those where the
gluon may be constrained from $F_2$ fits and jet production.}
\label{fig:plane}
\end{figure}

The kinematic plane at the LHC is shown in Fig.~\ref{fig:plane}, along
with the regions where LHC and other data will be able to constrain
the gluon density in the proton. There is an urgent need more
information about the gluon at high $x$ (say 0.05 and above) and at
$\q2$ between 100 and 10000 $\gevsq$, so that reliable predictions may
be made for the highest energy cross sections at LHC. In addition
there is a strong correlation between $\as$ and the gluon for
intermediate $x$ values (0.001 to 0.05) in fits to $F_2$.

Including DIS jet cross sections in the fit constrains the coupling,
but these cross sections are dominantly quark initiated and depend
only weakly on the gluon density.  Jet photoproduction, on the other
hand, is dominantly gluon initiated over a wide kinematic range, as
can reach very high $x$. ZEUS have included both in a
fit\cite{zeusfit}, with their latest inclusive cross section data, and
see a significant improvement in the accuracy of both $\as$ and the
gluon at high $x$. Perhaps most excitingly, the jet data used was a
fraction (around a tenth) of the total expected by the end of HERA
II. There are major improvements expected\cite{mcc}.

HERA II is also now producing high luminosities of electron-proton
collisions (rather than positron-proton), and early measurements were
shown at this conference. The large increase of statistics, matching
or bettering that achieved with positrons, and coupled with lepton
polarization, brings several benefits. One is the ability to measure
the electroweak structure of quark coupling (see a previous
contribution\cite{ew}). The measurement of charged and neutral
currents will also allow constraints on flavour composition of proton
to be made from HERA data alone, avoiding nuclear correction
uncertainties from fixed target data. These data also reach up to
high $x$.

\begin{figure}
\epsfxsize200pt
\figurebox{120pt}{170pt}{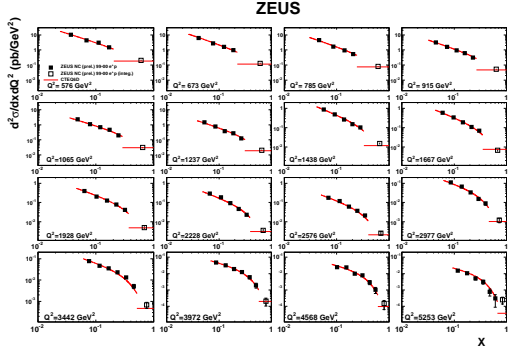}

\vspace{-0.5cm}

\caption{New measurements of $F_2$ at high $x$.}
\label{fig:highxf2}
\end{figure}

At lower $\q2$ it is still in principle possible to reach high $x$,
since the scattered electron may be measured. However, the radiative
corrections are such in this region that while reconstruction of $\q2$
from electron is good, it is very poor for $x$. A new measurement from
ZEUS\cite{zhix} uses the hadronic jet to reconstruct $x$. As $x$ increases, the
jet moves forward and will at some point be lost down the forward
beampipe. However, in this case it is possible to set a minimum $x$
based on the fact that the hadronic jet escaped, and integrate above
this. The measurement gives a good sensitivity to the high
$x$ structure function, as shown in Fig.~\ref{fig:highxf2}.

Finally in this subsection, the $W$ asymmetry measurements from
tevatron run II are now appearing\cite{wasym}. They are sensitive to flavour
composition in proton at high $x$ and will be important input to new
fits. 

\subsection{Low $Q^2$}

Measuring inclusive lepton-proton cross-sections in the low $\q2$
region probes the transition from a region where perturbative
calculations are valid to a region where non-perturbative techniques
must be used to make any prediction. It also provides the lowest reach
in $x$, and thus sensitivity to high density QCD. Two new measurements
from H1 have been presented in this area\cite{h1f2}. In the first,
QED Compton events, with a high virtuality exchanged electron, are
used. In this case the electron virtuality means that the final state
electron can be detected even when the virtuality of the exchanged
photon is very low. In the second such measurement, initial state
photon radiation is tagged, which implies a low virtuality incoming
electron with an energy lower than the beam energy. This incoming
electron energy is measured from the longitudinal energy imbalance in
the central detector. This allows the measurement to be made at lower
$\q2$ while keeping $x$ moderately high. Both of these measurements
provide new data in the transition region between DIS and
photoproduction.

\section{Peripheral Collisions, Low x and Diffraction}

The low $\q2$ region discussed above is an example of a measurement
where we deliberately extend into a region where the usual theoretical
tools are expected to fail. Moving into such regions allow the
investigation of new approximations in QCD such as clever
resummations, new evolution equations, new perturbative expansions,
high parton densities and correlated parton distributions. Using the
data to verify or falsify such tools extends our portfolio of
understood QCD phenomena. There is a large overlap in this area with
both the previous\cite{gavin} and following\cite{gunnar} speakers, and
I will concentrate on the topics least aligned with theirs.

\subsection{New resummations and evolutions}

\begin{figure}
\epsfxsize200pt
\figurebox{120pt}{170pt}{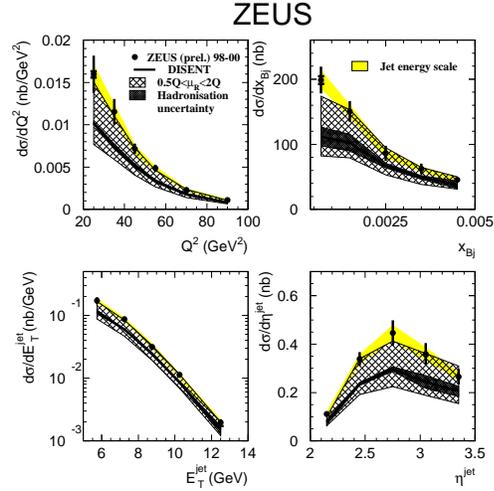}

\vspace{-0.5cm}

\caption{Forward jet cross sections at HERA.}
\label{fig:fw}
\end{figure}

The parton density fits discussed above all use the DGLAP evolution
equations, which are strongly ordered in the scale, $Q_1 \gg Q_2 \gg
Q_3$. For inclusive properties, this is the dominant
configuration. However, it is of course possible to select kinematic
configurations in which a large evolution in $x$ (or equivalently in
rapidity) is required, but where this evolution takes place at a $\q2$
which is both in the perturbative regime and approximately
constant. New measurements have been made in forward jet production
(Fig.\ref{fig:fw}) in DIS and other related processes at
HERA\cite{herafwj}.

In such a region the DGLAP evolution is not applicable. Thus if NLO
fixed order QCD with DGLAP parton densities is used to try and predict
such cross sections, the predictions have large uncertainties. It is
also seen that they usually lie below the data. Leading-logarithmic
Monte Carlos can do better than this, and in particular, the
CCFM-based MC \cascade\cite{cascade} probably has the ability to
describe such cross sections. However, it has a strong dependence on
the unintegrated gluon density, which is extracted from fits to
data. The new data should be used to constrain this further.

Such effects may also be studied in vector meson and photon
production.  The vector mesons I leave to the next
speaker\cite{gunnar}, but will mention here the new data from DELPHI
$\gamma^*\gamma^*$ collisions, where a signifcant $x$ evolution can
occur along the exchanged quark line. Again, calculations (BFKL-based)
which resum $\log(x)$ terms seem to have the best chance of describing
the data. 

A consistent, and reasonably precise, description of high rapidity/low
$x$ data seems to be within reach.  This would give a real boost to the
credibility of this approach, and would be a great help for predicting
forward jet rates at LHC.

\section{Conclusion}

In an increasing number of important processes at high energy
colliders, perturbative QCD calculations, and the data, are rather
precise, and in rather good agreement with each other.  New data from
Tevatron and HERA, and (re)analysis of old data from PETRA and LEP,
continue to improve the situation, as do theoretical advances. There
is still room for improvement of course, but for some important
processes QCD is now very precisely understood, and there have been
recent significant advances in measurement and theory. As an aside,
the point is now being reached where for some observables, electroweak
effects are comparable to QCD uncertainties\cite{ewcorr}. For other
processes, while QCD is becoming better understood, there is still
experimental and theoretical work to do. A list of such processes, in
approximate decreasing order of how well they are understood, could
be:

\begin{itemize}
\item Parton density functions at high $\q2$ and intermediate $x$, ideal jet
fragmentation.
\item Multijet processes, Boson+jets; Heavy flavour production.
\item Parton density functions at low and high $x$. 
\item High rapidities and rapidity gaps.
\item Diffraction, absorption and total cross sections.
\item Off-diagonal and unintegrated parton density functions.
\item Underlying events (a topic hardly touched on here, but where there is lots of work on tuning to Tevatron, HERA, SPS and other data\cite{heralhc,tev4lhc}).
\end{itemize}

In all these areas existing data, as well as data still to come from
Tevatron run II, HERA II and RHIC, provide a challenge. Data
from LHC will make great use of such developments, and will also
challenge the theory further.

\section*{Acknowledgments}

My thanks to D. Alton, R. Field, G. Ingelman, E. Perez, G. Salam,
J. Schiek, P. Wells and M. Wing for discussions and material, as well
as to those others whose work I have presented here.  I also thank the
organisers for a conference which was stimulating, well run and fun.

\end{document}